\documentclass[sigconf]{acmart}
\AtBeginDocument{%
  }

\copyrightyear{2026}
\acmYear{2026}
\setcopyright{cc}
\setcctype{by-nc-nd}
\acmConference[ICSE '26]{2026 IEEE/ACM 48th International Conference on Software Engineering}{April 12--18, 2026}{Rio de Janeiro, Brazil}
\acmBooktitle{2026 IEEE/ACM 48th International Conference on Software Engineering (ICSE '26), April 12--18, 2026, Rio de Janeiro, Brazil}
\acmPrice{}
\acmDOI{10.1145/3744916.3773164}
\acmISBN{979-8-4007-2025-3/2026/04}

\acmSubmissionID{icse2026-p2419}

\usepackage{listings}
\usepackage{enumitem}
\usepackage{amsfonts}
\usepackage{subcaption}
\usepackage{colortbl}
\usepackage{fancybox}
\usepackage{multirow}
\usepackage{fancybox}
\usepackage{xspace}
\usepackage{colortbl}
\usepackage{soul}
\usepackage{balance}

\newsavebox{\boxKbox}

\newenvironment{boxK}
{%
  \par\medskip\noindent%
  \setlength{\fboxsep}{10pt}
  \setlength{\fboxrule}{0.6pt}
  \begin{lrbox}{\boxKbox}%
    \begin{minipage}{\dimexpr\columnwidth-2\fboxsep-2\fboxrule\relax}%
      \small%
}
{%
    \end{minipage}%
  \end{lrbox}%
  \fcolorbox{black!25}{black!5}{\usebox{\boxKbox}}%
  \par\medskip%
}

\newcommand{\ie}{\textit{i.e.,}\xspace}
\newcommand{\eg}{\textit{e.g.,}\xspace}

\newcommand{\etal}{et al.\xspace}

\newcommand{\circled}[1]{\textcircled{\small #1}}

\newcommand\revision[1]{{#1}}

\newcommand{\equref}[1]{Eq.~\ref{#1}\xspace}
\newcommand{\secref}[1]{Sec.~\ref{#1}\xspace}
\newcommand{\figref}[1]{Fig.~\ref{#1}\xspace}
\newcommand{\tabref}[1]{Table~\ref{#1}\xspace}

\newcommand{\llm}{\textit{LLM}\xspace}
\newcommand{\llms}{\textit{LLMs}\xspace}

\newcommand{\psc}{\textit{PSC}\xspace}

\newcommand{\sept}{\textit{SECT}\xspace}
\newcommand{\ig}{\textit{IG}\xspace}

\definecolor{gradientplum}{RGB}{196,115,156}
\newcommand\codesmell[1]{\texttt{\textcolor{gradientplum}{\textbf{#1}}}}

\definecolor{gradientplum}{RGB}{196,115,156}
\newcommand\pycode[1]{\texttt{\textcolor{gradientplum}{\textbf{#1}}}}

\newcommand{\scm}{\textit{SCM}\xspace}

\newcommand{\ate}{\textit{ATE}\xspace}
\newcommand{\ates}{\textit{ATEs}\xspace}
\newcommand{\pearson}{$\rho$\xspace}

\newcommand{\anova}{\textbf{\textit{$F_{anova}$}\xspace}}
\newcommand{\pvalue}{\textbf{\textit{$p_{value}$}\xspace}}
\newcommand{\etasquared}{\textbf{\textit{$\eta^{2}$}\xspace}}
\newcommand{\confidence}{\textbf{\textit{$CI_{95\%}$}\xspace}}

\definecolor{gradientplum}{RGB}{196,115,156}
\newcommand\repository[1]{\textcolor{gradientplum}{\href{#1}{repository}}}

\begin{document}

\title{A Causal Perspective on Measuring, Explaining and Mitigating Smells in \llm-Generated Code}


\author{Alejandro Velasco}
\email{svelascodimate@wm.edu}
\orcid{0000-0002-4829-1017}
\affiliation{%
  \institution{William \& Mary}
  \city{Williamsburg}
  \state{Virginia}
  \country{USA}
}

\author{Daniel Rodriguez-Cardenas}
\email{dhrodriguezcar@wm.edu}
\orcid{0000-0002-3238-1229}
\affiliation{%
  \institution{William \& Mary}
  \city{Williamsburg}
  \state{Virginia}
  \country{USA}
}

\author{Dipin Khati}
\email{dkhati@wm.edu}
\orcid{0009-0008-4489-7733}
\affiliation{%
  \institution{William \& Mary}
  \city{Williamsburg}
  \state{Virginia}
  \country{USA}
}

\author{David N. Palacio}
\email{davidnad@microsoft.com}
\orcid{0000-0001-6166-7595}
\affiliation{%
  \institution{Microsoft}
  \city{Redmond}
  \state{Washington}
  \country{USA}
}

\author{Luftar Rahman Alif}
\email{bsse1120@iit.du.ac.bd}
\orcid{0009-0001-0092-6474}
\affiliation{%
  \institution{University of Dhaka}
  \country{Bangladesh}
  \city{Dhaka}
}

\author{Denys Poshyvanyk}
\email{dposhyvanyk@wm.edu}
\orcid{0000-0002-5626-7586}
\affiliation{%
  \institution{William \& Mary}
  \city{Williamsburg}
  \state{Virginia}
  \country{USA}
}

\renewcommand{\shortauthors}{Velasco et al.}

\begin{abstract}
\revision{Recent advances in large language models (\llms) have accelerated their adoption in software engineering contexts. However, concerns persist about the structural quality of the code they produce. In particular, \llms often replicate poor coding practices, introducing code smells (\ie patterns that hinder readability, maintainability, or design integrity). Although prior research has examined the detection or repair of smells, we still lack a clear understanding of how and when these issues emerge in generated code.}

\revision{This paper addresses this gap by systematically \textit{\textbf{measuring}}, \textit{\textbf{explaining}} and \textit{\textbf{mitigating}} smell propensity in \llm-generated code. We build on the Propensity Smelly Score (\psc), a probabilistic metric that estimates the likelihood of generating particular smell types, and establish its robustness as a signal of structural quality. Using \psc as an instrument for causal analysis, we identify how generation strategy, model size, model architecture and prompt formulation shape the structural properties of generated code. Our findings show that prompt design and architectural choices play a decisive role in smell propensity and motivate practical mitigation strategies that reduce its occurrence. A user study further demonstrates that \psc helps developers interpret model behavior and assess code quality, providing evidence that smell propensity signals can support human judgement. Taken together, our work lays the groundwork for integrating quality-aware assessments into the evaluation and deployment of \llms for code.}
\end{abstract}




\maketitle

\section{Introduction}
\label{sec:introduction}

\revision{\llms have rapidly moved from experimental tools to everyday assistants in software engineering (SE) \cite{watson2021systematicliteraturereviewuse}. Developers rely on LLM-based tools for many tasks such as code completion \cite{9616462, lin2024soen101codegenerationemulating, 7180092, sun2025bridgingdeveloperinstructionscode}, summarization \cite{ahmed2022fewshottrainingllmsprojectspecific, 11029737, 10.1145/3728949}, program repair \cite{jin2023inferfixendtoendprogramrepair, 10.1145/3650212.3680323}, clone detection \cite{10.1145/2970276.2970326, 10556419}, and test generation \cite{Watson_2020, 10329992, 10765033}. As generated code increasingly enters production systems, evaluation must look beyond functional correctness and consider whether the output meets broader standards of maintainable and robust software.}

\revision{Code smells provide a foundation for evaluating broader quality concerns by capturing recurring design and implementation issues that affect readability, complexity, and long-term maintainability \cite{smells_michele, palomba2018diffuseness, Evtikhiev_2023}. Prior work has traditionally attributed smell introduction to human-driven factors such as developer experience, time pressure, and project constraints \cite{smells_michele}. The growing involvement of \llms challenges this assumption. Models trained on large-scale code corpora inherit patterns present in the training data, including smelly or suboptimal code \cite{10795572, tambon2024bugslargelanguagemodels}, and may introduce additional defects or vulnerabilities during generation \cite{10.1145/3605098.3636058, pearce2021asleepkeyboardassessingsecurity}. Smell formation now reflects a combination of human design choices and model behavior, which renders traditional explanations incomplete.}


\revision{Current evaluation practices are not designed to reflect model decisions influencing smell generation. For instance, similarity metrics such as BLEU \cite{papineni2002bleu} and CodeBLEU \cite{codebleu}, along with execution-based measures, offer useful information about functional correctness but provide limited insight into structural or design quality \cite{empirical_code_smells, 10555682, nunes2025evaluatingeffectivenessllmsfixing, takerngsaksiri2025codereadabilityagelarge}. Although static analysis tools (SATs) can identify smells in model outputs, post-hoc detection alone falls short in revealing \textbf{why} a model introduced smells or \textbf{which} components of the generation process shaped them. As a result, developers lack visibility into factors that influence the maintainability of \llm-generated code.}

\revision{A deeper understanding of these issues requires analyzing how specific components of the generation pipeline influence both the \textit{structural and semantic properties} of the output. Viewed from this angle, several key questions emerge for the SE community: \textit{Which elements of the generation process affect the likelihood of smell introduction?} \textit{How do decoding strategies, model architectures, and prompt formulations shape the structure and maintainability of generated code?} \textit{Which of these factors can be intervened upon to guide models toward producing higher-quality outputs?}}

\revision{While these questions remain open, recent work has begun to introduce early indicators of smell propensity in \llms. Velasco \etal proposed the \psc metric, a probabilistic score that estimates the likelihood that a model generates tokens associated with a given smell \cite{velasco2025propense}. Introduced as part of a benchmarking effort, \psc provides a continuous signal of model tendencies and a promising foundation for investigating the factors behind smell occurrence. Yet the broader dynamics of smell generation remain unexplored. 1) The drivers of variation in \psc and their relationship to the structural quality of generated code are still unknown, and 2) the stability of the metric across diverse conditions remains untested. Finally, 3) \psc has yet to be used to uncover mechanisms that explain \emph{why} models introduce smells or to support interventions that reduce smell propensity. These three gaps hinder a full understanding of how smells emerge in \llm-generated code and limit our ability to devise approaches for reducing their occurrence.}


\revision{Responding to these gaps, we investigate smell generation from a causal perspective. Instead of treating \psc as a standalone benchmark metric, we use it as an instrument for reasoning about the factors that influence smell propensity in \llm-generated code. Our study begins by assessing the robustness and explanatory value of \psc, establishing that it provides a stable foundation for causal analysis. We then quantify the effects of key elements in the generation process, including decoding strategy, model size, model architecture, and prompt formulation, on the likelihood of smell occurrence. These causal analyses identify the factors that meaningfully \textit{influence smell propensity} and guide the design of \textit{mitigation strategies} that can be applied during inference. Finally, we complement this analysis with a user study that evaluates whether \psc helps developers interpret and assess generated code.}

\revision{Our results show that prompt formulation and model architecture most strongly influence smell propensity, and that targeted adjustments to these factors can significantly reduce smell introduction during inference. We also find that \psc captures structural quality signals overlooked by traditional metrics and supports developers in reasoning about generated code. Our specific contributions are listed as follows:}

\revision{
\begin{itemize}
    \item We validate \psc as a stable and informative instrument for analyzing smell propensity.
    \item We provide a causal analysis quantifying the influence of generation strategy, model size, model architecture, and prompt formulation.
    \item We design prompt-based mitigation strategies that reduce smell propensity during inference.
    \item We conduct a user study evaluating the practical value of \psc for developers.
    \item We release a dataset, benchmark, and implementation to support future research on explaining and mitigating smell generation \cite{psc_repo}.
\end{itemize}
}

\section{Motivation \& Related Work}
\label{sec:motivation_related_work}

\textit{Limitations of current evaluations.} Existing evaluation methods prioritize surface-level correctness or lexical similarity. As a result, outputs that pass canonical metrics may still exhibit design flaws or maintenance issues. Recent empirical studies have highlighted this gap. Siddiq \etal \cite{empirical_code_smells} and Torek \etal \cite{melin2024precisionperilevaluatingcode} showed that both the training data and the code generated from \llms frequently exhibit code smells and insecure constructs. Alif \etal \cite{mohsin2024trustlargelanguagemodels} observed that models like Bard and ChatGPT introduce more smells than Copilot or CodeWhisperer in complex tasks. Similarly, Kharma \etal \cite{kharma2025securityqualityllmgeneratedcode} linked outdated training corpora to the persistence of legacy antipatterns. In the context of test generation, Ouédraogo \etal \cite{ouedraogo2024testsmellsllmgeneratedunit} found high rates of test-specific smells in the outputs of GPT-3.5 and GPT-4. Although these findings underscore the need for deeper evaluation, current tools do not provide standardized methods to measure or explain the structural quality of \llm-generated code.


\textit{Toward principled evaluation and understanding.} To address these limitations, recent work has introduced specialized benchmarks and metrics aimed at evaluating and explaining the structural quality of \llm-generated code. The \textit{CodeSmellEval} benchmark \cite{velasco2025propense} proposed the Propensity Smelly Score (\psc), a probabilistic metric that estimates the likelihood of generating specific types of smell. Zheng \etal \cite{zheng2024correctnessbenchmarkingmultidimensionalcode} developed the RACE framework to incorporate broader quality dimensions such as maintainability and readability. Wu \etal \cite{wu2024ismell} introduced \textit{iSMELL}, a system that combines static analysis with \llms to improve smell detection and refactoring. Although these approaches mark important progress, key questions remain unanswered: \textit{Which generation factors most influence the presence of smell? Can we isolate causal contributors to quality degradation? And can smell generation be systematically reduced through actionable strategies?}


This paper builds upon prior work by shifting the focus from merely detection to explanation and mitigation. We aim to systematically \textbf{measure}, \textbf{explain}, and \textbf{mitigate} code smells in \llm-generated code. To support this goal, we adopt the \psc as a measurement tool and extend its use across three dimensions.

\section{Propensity Smelly Score (\psc)}
\label{sec:background}
\begin{figure}[ht]
		\centering
  \includegraphics[width=0.35\textwidth]{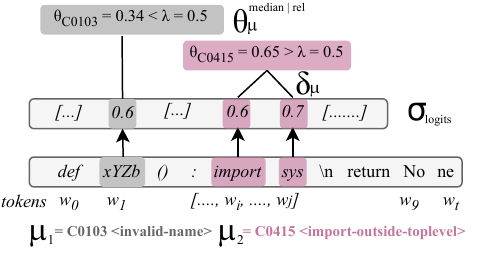}
		\caption{Propensity Smelly Score (\scm) Computation. The python snippet at the bottom contains two code smells: \codesmell{C0103} (\ie invalid-name) and \codesmell{C0415} (\ie import-outside-toplevel).}
    \label{fig:psc_computation}
\end{figure}

The Propensity Smelly Score (\psc) is a probabilistic metric that estimates the likelihood that a \llm generates a specific type of code smell \cite{velasco2025propense}. It relies on the next-token probabilities predicted by the model during autoregressive generation. Given a token sequence $w = \{w_1, w_2, \ldots, w_n\}$, the model produces a probability distribution in the vocabulary for each token position. The \psc computation process is depicted in \figref{fig:psc_computation}.

Let $\mu$ denote a code smell instance aligned to a token span $(i, j)$ in $w$ using an alignment function $\delta_\mu(w)$. The \psc is computed by aggregating the model's token-level probabilities within this span. For each token $w_k$ in the range $i \leq k \leq j$, we compute $P(w_k \mid w_1, \ldots, w_{k-1})$ using softmax-transformed logits. The aggregated \psc for $\mu$ is given by Equation~\ref{eq:aggregation_function}:

\begin{equation}
\label{eq:aggregation_function}
\theta_\mu(w, i, j) = \mathbb{E}_{k = i}^{j} \left[ P(w_k \mid w_1, \ldots, w_{k-1}) \right]
\end{equation}


The function $\theta_\mu$ reflects the model's average confidence in generating the sequence associated with a given code smell. Higher \psc values indicate a stronger tendency to produce tokens that correspond to known smells. \revision{To support comparisons across models, we adopt a threshold $\lambda = 0.5$ from prior work \cite{velasco2025propense} as a default interpretability criterion, where \textit{a \psc value at or above $\lambda$ indicates that the model is likely to generate the corresponding smell}. This threshold is used only to interpret the score and does not affect how \psc is computed. Adjusting it changes how many instances are labeled as propense but leaves the underlying probabilities unchanged.}

\subsection{Computation of the Aggregation Function $\theta$}

The aggregation function $\theta_\mu(w, i, j)$ (\equref{eq:aggregation_function}) supports multiple computation strategies to estimate \psc. Common approaches include taking the mean or median of token-level probabilities within the span associated with the smell. Alternatively, relative aggregation methods can be applied to normalize probabilities based on the surrounding context or the baseline distributions.

\paragraph{Mean and Median Aggregation.}
The simplest approach aggregates the actual probabilities $P(w_k \mid w_1, \ldots, w_{k-1})$ over the token span $(i, j)$ Two common options are the mean, defined as $\theta_\mu^{\text{mean}} = \frac{1}{j - i + 1} \sum_{k = i}^{j} P(w_k \mid w_1, \ldots, w_{k-1})$, and the median, defined as $\theta_\mu^{\text{median}} = \text{median} \left( \{ P(w_k \mid w_1, \ldots, w_{k-1}) \mid i \leq k \leq j \} \right)$. The mean captures the overall trend in model confidence but may be influenced by outliers, whereas the median is more robust to local fluctuations and low-confidence predictions.

\textit{Relative Aggregation.} To account for variability across code smells or token positions, we introduce a normalized variant of the score, called the \textit{relative propensity score}. This version rescales each token probability using empirical bounds observed in a reference dataset:

\begin{equation}
\label{eq:psc_relative}
\theta_\mu^{\text{rel}}(w, i, j) = \frac{1}{j - i + 1} \sum_{k = i}^{j} \frac{P(w_k) - P_{\text{min}}(w_k)}{P_{\text{max}}(w_k) - P_{\text{min}}(w_k) + \epsilon}
\end{equation}

Here, $P(w_k)$ denotes the actual probability for token $w_k$, while $P_{\text{min}}(w_k)$ and $P_{\text{max}}(w_k)$ represent the empirical minimum and maximum values for the same token position within the sample. A small constant $\epsilon$ is added to ensure numerical stability. This formulation scales all values to a unit range, facilitating fair comparisons across smells and models.

\textit{Implementation}. In practice, all three variants of $\theta$ are computed from the same set of token-level probabilities. Each reflects a distinct aggregation philosophy: the mean and median quantify absolute confidence, while the relative score captures scale-invariant trends that are useful for comparative evaluations.

\section{Methodology}
\label{sec:methodology}



\revision{This section presents our methodology for addressing four research questions on measuring, explaining and mitigating code smells in \llm-generated code, as well as evaluating the practical utility of \psc. We begin by validating \psc through robustness and information-gain analyses. We then apply causal inference to identify the factors that shape smell propensity. Building on these insights, we design a mitigation case study and conclude with a user study examining how practitioners interpret and apply \psc during code review.}

\begin{enumerate}[label=\textbf{RQ$_{\arabic*}$}, ref=\textbf{RQ$_{\arabic*}$}, wide, labelindent=5pt]\setlength{\itemsep}{0.2em}
     \item \label{rq:measure} {\textbf{[Measure]} \textit{How can we measure the propensity of LLMs to generate code smells?}} We rely on \psc as a probabilistic indicator of smell propensity. To assess its effectiveness, we examine its \textit{robustness} under \sept and compare its \textit{information gain} with baselines such as BLEU and CodeBLEU.

      \item \label{rq:explain} {\textbf{[Explain]} \textit{What factors contribute to the presence of code smells in \llm-generated code?}} We examine four intervention scenarios on the behavior of \psc: \textit{model architecture}, \textit{model size}, \textit{generation strategy} and \textit{prompt type}. Using causal inference techniques, we estimate the effect of each intervention on the score.

      \item \label{rq:mitigate} {\textbf{[Mitigate]} \textit{What strategies can reduce the presence of code smells in code generated \llm?}} Based on the insights obtained from \ref{rq:explain}, we conduct a case study to demonstrate how controlling these factors can reduce \psc, thus lowering the model's tendency to produce smelly code.
      
     \item \label{rq:usefulness} {\textbf{[Usefulness]} \textit{How do developers assess code smells in \llm-generated code when supported by \psc?}} We conduct a user study to examine whether \psc helps developers evaluate generated code, focusing on its perceived relevance, its impact on confidence, and its usefulness in identifying introduced smells.

\end{enumerate} 

\subsection{Semantic Preserving Code Transformations}

To assess the robustness of the \psc metric (\ie \ref{rq:measure}) under syntactic variation, we design our methodology based on \textit{Semantic Preserving Code Transformations} (\sept). The goal is to evaluate whether \psc remains consistent when the input code is modified in ways that preserve functionality but alter its syntax. A reliable quality metric should remain stable under such modifications, as the semantic behavior of the code remains unchanged.

\revision{Our methodology draws on prior work in robustness evaluation for neural program repair using natural transformations \cite{sept_paper}. We focus specifically on \textbf{statement-level transformations}. Broader edits, such as converting while loops to for loops (\textit{While2For}), can significantly alter the code's structural context, thereby influencing the detection of code smells. Limiting the analysis to statement-level changes preserves semantics and overall structure while still introducing meaningful syntactic variation.}



\revision{We apply a curated set of six transformations, listed in \tabref{tab:sept}, including \textit{Add2Equal}, \textit{SwitchEqualExp}, \textit{InfixDividing}, \textit{SwitchRelation}, \textit{RenameVariable-1} and \textit{RenameVariable-2}. These transformations modify the lexical and syntactic form of code statements without altering control flow or observable behavior. To evaluate robustness, we compute the \psc using the \textbf{relative aggregation} method (see \equref{eq:psc_relative}) for code snippets containing confirmed instances of each smell type. Because raw \psc values are bounded and often skewed, we apply a \textbf{logit transformation} to improve normality prior to statistical testing.}




\revision{We conduct a one-way \textbf{Analysis of Variance (ANOVA)} to test whether \psc scores differ across transformation types, performing this analysis separately for each smell. The results indicate where the metric remains stable under syntactic variation and where it becomes more sensitive to surface-level changes. For each case, we report the $F$-statistic (\anova), the $p$-value (\pvalue) and the effect size (\etasquared), where non-significant $p$-values and small effects indicate robustness under syntactic changes. To complement these tests, we compute 95\% confidence intervals (\confidence) for relative \psc scores across transformation variants. Lower variance across intervals suggests stability under syntactic changes, whereas higher variance indicates susceptibility to perturbations.}

\subsection{Information Gain (\ig)}
\label{sec:ig}

To assess how well different metrics explain the presence of severe code smells in generated code (\ie \ref{rq:measure}), we calculate the information gain (\ig) between each metric and a binary severity label. This enables a direct comparison between \psc, which is based on token-level generation probabilities, and existing baselines such as BLEU and CodeBLEU, which rely on similarity to a reference output.

\revision{BLEU and CodeBLEU are widely used in code generation benchmarks and are often treated as default baselines. Both rely on surface-level similarity, using n-gram overlap and syntax-based matching to assess how closely a generated snippet aligns with a reference. These metrics are effective for evaluating correctness but are not designed to reflect deeper structural properties such as code smells. Because no established alternatives exist for assessing structural quality in generated code, we include BLEU and CodeBLEU as representative standards for comparison \cite{Evtikhiev_2023, melin2024precisionperilevaluatingcode}. Their reliance on reference similarity contrasts with \psc, which captures likelihood-based signals from the model itself. To compare metrics that draw on such different sources of information, we adopt an information-gain framework that quantifies how much each metric reduces uncertainty about the severity of code smells.}

To define severity, we analyze the proportion of smelly tokens in a generated snippet. A snippet is labeled as \textit{high severity} if more than fifty percent of its tokens are flagged as smelly based on static analysis. Otherwise, it is labeled as \textit{low severity}. Let $n_s$ be the number of smelly tokens and $n_t$ the total number of tokens in the snippet. The severity label is assigned according to the following rule:

\begin{equation}
    S = 
    \begin{cases}
        \text{high}, & \text{if } \frac{n_s}{n_t} > 0.5 \\
        \text{low}, & \text{otherwise}
    \end{cases}
\end{equation}

Let $S \in \{\text{low}, \text{high}\}$ denote the severity label and let $X$ be a continuous score produced by a given metric, such as \psc, BLEU or CodeBLEU. The Information Gain is defined as:

\begin{equation}
    \text{IG}(S, X) = H(S) - H(S \mid X)
\end{equation}

Here, $H(S)$ is the entropy of the severity distribution, and $H(S \mid X)$ is the conditional entropy of $S$ given the score $X$. A lower conditional entropy indicates that the metric provides more information about severity by reducing uncertainty. Therefore, a higher information gain value reflects a stronger relationship between the metric and structural quality.

We expect \psc to produce a higher information gain than BLEU and CodeBLEU. Because \psc directly captures the model's generation behavior with respect to smelly structures, it should align more closely with severity labels derived from token-level analysis. In contrast, BLEU and CodeBLEU may not reflect such quality differences, especially in cases where syntactically similar outputs differ in maintainability. A higher information gain for \psc would confirm its suitability for evaluating design-level concerns in generated code, such as code smells.

\subsection{Causal Analysis of Factors Influencing \psc}

\begin{figure}[ht]
		\centering
  \includegraphics[width=0.4\textwidth]{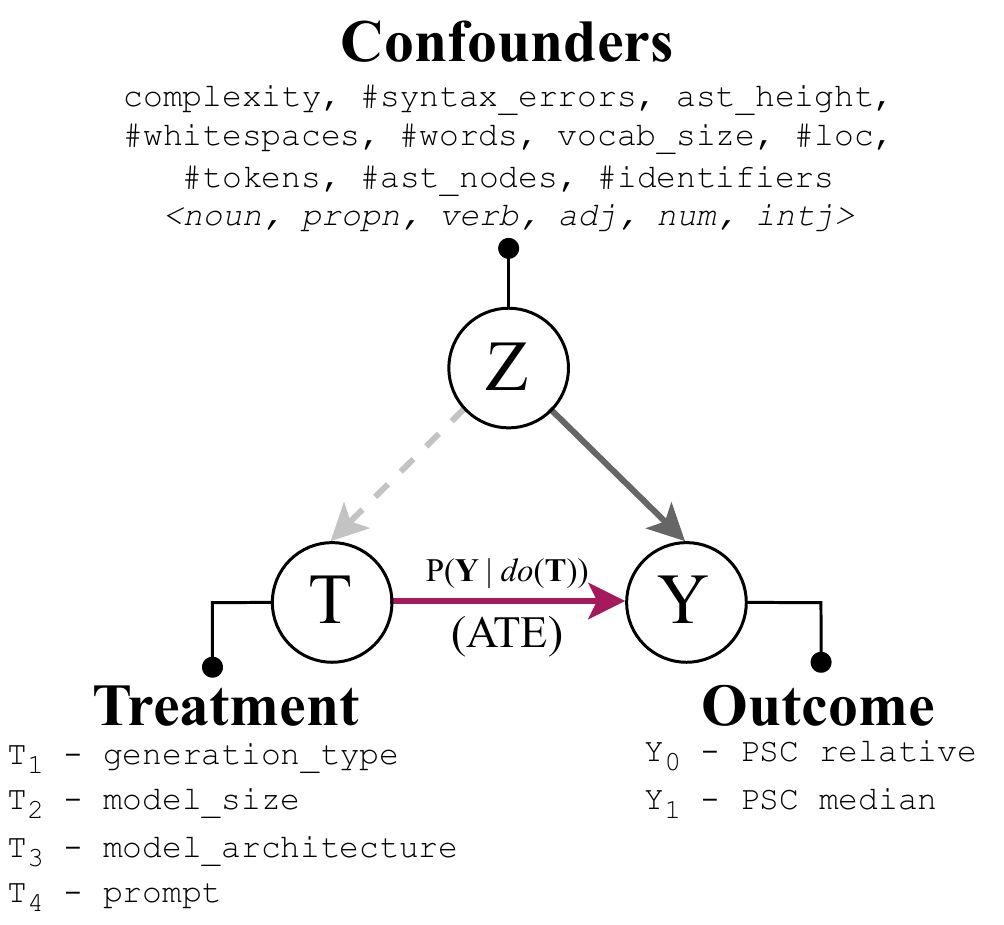}
		\caption{Structural Causal Model (\scm)}
    \label{fig:scm}
\end{figure}

To quantify the causal effect of the intervention settings on \psc (see \ref{rq:explain}), we construct a structured causal model (\scm), illustrated in \figref{fig:scm}. Our objective is to estimate the effect of four treatment variables: generation strategy ($T_1$), model size ($T_2$), model architecture ($T_3$), and prompt type ($T_4$), on two outcomes: relative \psc ($Y_0$) and median \psc ($Y_1$).

\revision{We adjust for syntactic and lexical features that confound the relationship between treatments and outcomes (\ie $Z$). These include lines of code (LOC), token count, number of AST nodes and identifiers, AST height, number of syntax errors and whitespaces, word count and vocabulary size. We also control for the distribution of common part-of-speech (POS) tags in code, such as number of nouns, proper nouns, verbs, adjectives, numerals and interjections.}

\revision{Previous studies support the inclusion of these variables. Sampling strategies with high temperature tend to produce longer and more deeply nested code, whereas greedy decoding yields shorter and flatter sequences~\cite{donato2025studyingconfigurationsimpactcode}. Model size shapes output as well. Larger models typically produce more concise and focused completions, while smaller ones often generate repetitive or verbose code~\cite{chen2021evaluatinglargelanguagemodels}. Architectural design introduces further variation. Models that use grouped-query attention and sliding windows, such as Mistral, often produce shorter sequences than similarly sized LLaMA models~\cite{jiang2023mistral7b}. Prompt formulation has a comparable effect. Detailed prompts, including full function signatures, tend to elicit longer and more structured outputs, whereas vague prompts result in shorter completions~\cite{khojah2024impactpromptprogrammingfunctionlevel}. Finally, probabilistic metrics reflect the underlying structure of code. The ASTTrust study~\cite{palacio2024syntax} shows that metrics such as \psc vary with AST complexity and token alignment.}

\subsubsection{Computing Average Treatment Effects}

\revision{The average treatment effect (\ate), shown in \figref{fig:scm}, captures the expected change in \psc resulting from an intervention, averaged over the distribution of confounders. Formally, \ate is expressed as $P(Y \mid \text{do}(T))$ in Pearl's do-calculus~\cite{Pearl2018Causality}. We identify valid causal estimands by applying graphical criteria (\eg back-door criterion), which determines the appropriate adjustment set $Z$ from the causal graph.}

\subsubsection{Measuring Robustness}


\revision{To assess the reliability of the estimated \ate values, we apply four refutation tests. We add random common causes to check for spurious effects, run placebo tests by replacing the treatment with random variables, simulate unobserved confounding to gauge sensitivity to missing factors and perform subset validation to examine stability across data partitions. Consistent patterns across these tests indicate that the estimates are robust and generalizable.}

\subsubsection{Interpreting Average Treatment Effects}

A positive \ate indicates that treatment increases \psc, suggesting a higher likelihood of generating code smells. In contrast, a negative \ate implies a decrease in \psc, indicating a potential reduction in the generation of smells. The magnitude of \ate reflects the strength of the causal effect. Its credibility is supported by statistical significance and robustness to sensitivity tests.

\subsection{\psc Mitigation}


To address \ref{rq:mitigate}, we investigate whether prompt design can mitigate the generation of code smells as measured by \psc. Instead of modifying model parameters or architecture, we focus on inference-time interventions that remain accessible and practical for users.

\revision{Our case study builds on the findings of \ref{rq:explain} (specifically $T_4$), where prompt type emerged as a significant causal factor influencing \psc. To operationalize this insight, we isolate the effect of prompt formulation by comparing two strategies: a minimal prompt that presents only a Python function containing a smell, and a structured prompt that provides an explicit task description.}

For each smell instance, we generate completions under both prompt conditions using the same underlying model. We then compute the \psc for the smelly portion of each generated output and compare the distributions between conditions to assess the impact of prompt design on code smell propensity.

\subsection{\psc Usefulness} 

To address \ref{rq:usefulness}, we conducted a user study between subjects in order to evaluate the practical value and perceived usefulness of the \psc metric when integrated in the context of code review. This section describes the study design in detail. We also report on how the instrument was validated prior to deployment to ensure clarity and usability.

\subsubsection{Population Profiling} The target population for this study consisted of people with experience in software development and familiarity with code generation tools powered by \llms, such as \textit{GitHub Copilot} and \textit{ChatGPT}. Participants were expected to have working knowledge of Python and a general understanding of programming concepts relevant to code structure and maintainability, including the notion of code smells. Although experience with deep learning models (\eg GPT, T5, or LLaMA) was considered beneficial, it was not required. Our goal was to capture perspectives from academic and industry professionals with practical exposure to \llm-assisted programming. Participants were required to be at least 21 years of age. We did not impose restrictions on gender, educational level, or professional background beyond basic software development proficiency. Participation was voluntary and individuals were informed about the purpose of the study and their right to withdraw at any time.

\subsubsection{Sampling} We followed a \textit{ purpose sampling} strategy \cite{sampling}, since our objective was to collect early evidence and practitioner feedback on the usefulness of \psc to assess the propensity to generate code smells at the snippet level. The participants were selected based on their combined experience in machine learning and software engineering. We prioritized relevant expertise over randomness to ensure a meaningful interpretation of the metric output. Invitations were sent by email across academic and professional networks, and recipients were encouraged to share the survey with peers.

\subsubsection{Data Collection} We contacted approximately $110$ potential participants from academic and industry settings, in order to collect perspectives from individuals with varying levels of experience in software engineering. Invitations were sent by email and participants were not informed of the specific goals of the study during recruitment. The survey was administered using Qualtrics. In total, we received 49 responses: 25 for the control survey and 24 for the treatment survey. After filtering out incomplete or low-quality submissions, such as those with unanswered questions or only demographic information, we retained 36 valid responses. The final dataset consisted of 18 control responses and 18 treatment responses.

\subsubsection{Survey Structure} The survey consisted of three sections: background profiling, code evaluation, and post-task reflection. Both the control and treatment groups followed the same structure, with the only distinction being the inclusion of the \psc score and two additional questions in the treatment. The survey began by collecting background information such as the current role of the participants, years of programming experience, familiarity with Python and code smells, and prior use of LLMs in software development workflows. In the second section, five Python code snippets were shown to the participants, each associated with a specific code smell detected using \texttt{Pylint}. The control group viewed the code along with a textual description of the smell, while the treatment group also saw the corresponding \psc score. For each snippet, all participants answered three questions: ($Q_1$) \textit{“How likely do you think this code smell was introduced systematically or intentionally by the model (rather than being accidental)?”}; ($Q_2$) \textit{“How important do you think it is to review or fix this code smell in a real-world codebase, considering the potential risk of similar issues reappearing in the future?”}; and ($Q_3$) \textit{“How confident are you in your judgment about this code smell?”} The treatment group received two additional questions: ($Q_4$) \textit{“Did the \psc score influence your judgment about this code smell?”} and ($Q_5$) \textit{“Please briefly explain how the \psc score influenced your decision.”} The final section of the survey asked participants to reflect on their general trust in \llm-generated code and their interest in integrating automated smell detection tools into development workflows.

\subsubsection{Statistical Analysis} To evaluate the impact of \psc exposure on participant responses, we applied non-parametric statistical tests suitable for ordinal data. Specifically, we used the Mann–Whitney U test to compare responses between the control and treatment groups for each of the three main questions ($Q_1$–$Q_3$). This test does not assume normality and is appropriate for detecting differences in central tendency between independent samples with Likert scale results. Statistical comparisons were performed independently for each question and code smell type, yielding a total of 15 comparisons (3 questions in 5 snippets). For each comparison, we report the $p$ value and highlight statistically significant differences ($p < .05$). \revision{Open-ended responses from $Q_5$ were analyzed using thematic analysis \cite{themes}, with two authors independently coding answers and consolidating recurring ideas into broader themes.}

\subsubsection{Survey Validity}  
We validated the survey through a pilot study with four doctoral students experienced in SE and familiar with machine learning models. None of them were involved in this work. They completed both control and treatment versions and provided feedback on question clarity, code readability and the interpretability of the \psc explanation. Their input led to revisions in prompt phrasing, improvements to the survey layout and refinements to the wording used to describe \psc.

\section{Experimental Context}

This section outlines the empirical setting for our study and details the data, models, and experimental conditions used to address our research questions. We begin by describing the construction of the data set used for analysis. We then present the models and treatment interventions evaluated throughout the quantification, explanation, mitigation, and usability tasks.

\label{sec:experimental_context}
\begin{table}[]

\centering
\caption{\llms used in experiments.}
\label{tab:models}
\scalebox{0.77}{

\setlength{\tabcolsep}{4pt} 
\begin{tabular}{lllllll}
\multicolumn{3}{c}{\textit{Same Size}}       &           & \multicolumn{3}{c}{\textit{Same Architecture}} \\
\textbf{ID} & \textbf{Model} & \textbf{Size} & \textbf{} & \textbf{ID}  & \textbf{Model}  & \textbf{Size} \\ \hline
$M_1$ & CodeLlama-7b-hf \cite{codeLlama} & 7B &  & $S_1$ & Qwen2.5-Coder-0.5B \cite{qwen} & 0.5B \\
$M_2$ & Mistral-7B-v0.3 \cite{mistral}  & 7B &  & $S_2$ & Qwen2.5-Coder-1.5B \cite{qwen} & 1.5B \\
$M_3$ & Qwen2.5-Coder-7B \cite{qwen} & 7B &  & $S_3$ & Qwen2.5-Coder-3B \cite{qwen}   & 3B   \\
$M_4$ & starcoder2-7b \cite{starcoder}   & 7B &  & $S_4$ & Qwen2.5-Coder-7B \cite{qwen}   & 7B  
\end{tabular}
} 

\end{table}
\begin{table}[]

\centering
\caption{List of \sept}
\label{tab:sept}
\scalebox{0.7}{

\setlength{\tabcolsep}{4pt} 

\begin{tabular}{llll}
\textbf{Transformation} &
  \textbf{Description} &
  \textbf{Original} &
  \textbf{Transformed} \\ \hline
\textit{Add2Equal} &
  \begin{tabular}[c]{@{}l@{}}Convert add/subtract assignments\\ to equal assignments\end{tabular} &
  \pycode{a += 9} &
  \pycode{a = a + 9} \\
\textit{SwitchEqualExp} &
  \begin{tabular}[c]{@{}l@{}}Switch the two expressions on \\ both sides of the infix expression \\ whose operator is ==\end{tabular} &
  \pycode{a == b} &
  \pycode{b == a} \\
\multirow{2}{*}{\textit{InfixDividing}} &
  \multirow{2}{*}{\begin{tabular}[c]{@{}l@{}}Divide an in/pre/post-fix \\ expression into two expressions\end{tabular}} &
  \multirow{2}{*}{\pycode{x = a + b * c}} &
  \multirow{2}{*}{\begin{tabular}[c]{@{}l@{}}\pycode{temp = b *c} \\ \pycode{x = a + temp}\end{tabular}} \\
 &
   &
   &
   \\
\textit{SwitchRelation} &
  Transform relational expressions &
  \pycode{a \textgreater b} &
  \pycode{b \textless a} \\
\textit{RenameVariable-1} &
  \begin{tabular}[c]{@{}l@{}}Replace a variable name by its \\ first character\end{tabular} &
  \pycode{number = 1} &
  \pycode{n = 1} \\
\textit{RenameVariable-2} &
  \begin{tabular}[c]{@{}l@{}}Replace a variable name by \\ substitutions from CodeBERT\end{tabular} &
  \pycode{number = 1} &
  \pycode{myNumber = 1}
\end{tabular}
} 
\end{table}

\subsection{Dataset}
\label{sec:dataset}

Our experiments are based on an expanded version of the \textit{CodeSmellData} corpus \cite{velasco2025propense}, referred to as \textit{CodeSmellData 2.0}. This dataset was constructed through a multistage pipeline designed to extract and annotate large-scale Python code snippets from open-source repositories, allowing empirical evaluation of code smell generation in \llm-generated code.

\textit{Data Mining and Processing.} We began by extracting method-level Python snippets from public GitHub repositories with more than $1K$ stars, published between January 2020 and January 2025. This process yielded approximately $4.7M$ methods, which were stored in a MySQL database. Each snippet was processed to compute syntactic and lexical confounders using the GALERAS schema \cite{galeras}. In addition to these features, we extracted part-of-speech (POS) token counts using the \texttt{spaCy} parser to support fine-grained structural and linguistic analysis.

\textit{Deduplication and Annotation.} We removed exact and semantically similar duplicates by comparing abstract syntax tree (AST) representations, resulting in a deduplicated set of approximately $2.1M$ unique methods exported to JSON format. Methods that did not perform the AST parsing were excluded. The remaining snippets were analyzed using \texttt{Pylint} to identify code smells, recording their types and sources. This step produced around $2.8M$ unique code smell instances, which collectively comprise \textit{CodeSmellData 2.0} \cite{psc_repo}.

\textit{Filtering for Experimental Use.} Due to computational constraints and the need for uniform sampling, we applied two additional filters to produce the dataset used in our experiments. First, we excluded any snippet that exceeded $700$ tokens when tokenized using the \texttt{CodeLlama-7b-hf} ($M_1$) tokenizer. Second, for each type of code smell, we retained exactly $500$ randomly sampled instances. Smell types with fewer than $500$ examples were excluded from the analysis. The resulting evaluation dataset includes $41$ distinct smell types covering the \textbf{warning}, \textbf{convention}, and \textbf{refactor} categories defined by \texttt{Pylint}.

\tabref{tab:robustness_results} lists the included smell types and their \texttt{Pylint} identifiers. Each type is capped at 500 instances for consistency. However, the full distribution of smells from \textit{CodeSmellData 2.0} is preserved in the dataset released \cite{psc_repo}.

\subsection{Models}

We selected pre-trained decoder-only models that use causal autoregressive decoding, making them well suited for the token-level probability computations required by \psc. These architectures are standard in code generation due to their ability to model long-range dependencies. The chosen models support controlled comparisons across architecture and scale. \tabref{tab:models} summarizes the full set of selected models.

Although our implementation of \psc is designed for autoregressive decoding with access to token probabilities, the method is not limited to this setting. The metric can be extended to other types of generative architecture, including encoder-only or encoder-decoder models, provided that token-level likelihoods are accessible during inference.


\subsection{Causal Settings}
\label{sec:causal_settings}

\begin{figure}[ht]
		\centering
  \includegraphics[width=0.48\textwidth]{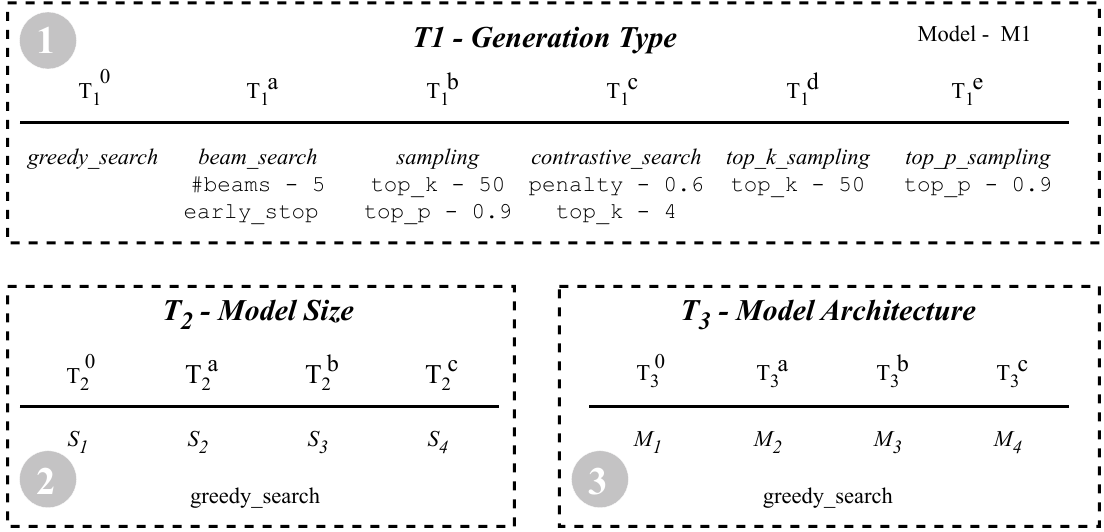}
		\caption{Causal Analysis $T_{1-3}$ Settings}
    \label{fig:causal_settings_1}
\end{figure}

Each intervention in $T$ (refer to \figref{fig:scm}) is designed to evaluate the influence of specific factors that we hypothesize contribute to \psc. These factors include: $T_1$, the type of generation; $T_2$, the model size; $T_3$, the model architecture (depicted in \figref{fig:causal_settings_1}); and $T_4$, the type of prompt (depicted in \figref{fig:causal_settings_2}). We define the interventions for each treatment as follows.

\subsubsection{Intervention Setting $P_{}(Y_1|do(T_1)$}

The $T_1$ intervention (see \circled{1} in \figref{fig:causal_settings_1}) manipulates the generation strategy used by the language model during inference. We evaluated six decoding methods: \textit{greedy search}, \textit{beam search}, \textit{sampling}, \textit{contrastive search}, \textit{top-$k$ sampling}, and \textit{top-$p$ sampling}. The beam search is configured with five beams and early stopping enabled. Contrastive search uses a penalty of 0.6 and a top-$k$ value of 4. The top$k$ sampling is applied with $k=50$, and the top$p$ sampling uses $p=0.9$. All decoding strategies are applied to the same model instance ($M_1$) to ensure comparability between conditions.

For the code snippet in the dataset (\secref{sec:dataset}), we truncate 50\% of its original token length and instruct the model to complete the remaining portion using a specific decoding strategy. This 50\% cutoff provides a consistent balance between context and generation length across examples. We then apply \texttt{Pylint} to the completed full snippet to detect any code smells. Although it is possible that the prefix may already contain smells, we analyze only those that appear in the generated portion. Smells are retained for analysis only if their source location falls entirely within the model-generated segment. For each of these, we compute the corresponding \psc ($Y_1 = f(\theta_\mu^{\text{median}} )$) to quantify the propensity of the model to produce the detected issue.

We define \textit{greedy search} as the control condition, denoted by $T_1^0$. Each alternative decoding strategy serves as a separate treatment condition, denoted $T_1^{a}$ through $T_1^{e}$. We estimate average treatment effects (\ates) by comparing each $T_1^{a-e}$ setting against the control.

\subsubsection{Intervention Setting $P(Y_1 \mid do(T_2))$}

The $T_2$ intervention (see \circled{2} in \figref{fig:causal_settings_1}) examines the effect of model size on code smell propensity. We evaluated four variants of the \texttt{Qwen2.5-Coder} model family, ranging from $0.5B$ to $7B $ parameters. These are denoted $S_1$ through $S_4$ in \tabref{tab:models}. To isolate the effect of model size, all models share the same architecture and are evaluated under a fixed decoding strategy (\textit{greedy search}).

For each model, we compute the \psc ($Y_1 = f(\theta_\mu^{\text{median}} )$) for every code snippet that contains a smell instance in the evaluation dataset, treating the snippet as a fixed input rather than generating it from scratch. Using an identical code across models controls the variability of generation and isolates the causal effect of the size of the model on \psc.

In the causal setting, the smallest model $S_1$ serves as the control condition, denoted $T_2^0$. The remaining models $S_2$, $S_3$, and $S_4$ are treated as treatment conditions, labeled $T_2^{a}$, $T_2^{b}$, and $T_2^{c}$, respectively. We estimate average treatment effects by comparing the distribution of $Y_1$ across these size-based configurations.

\subsubsection{Intervention Setting $P(Y_1 \mid do(T_3))$}

The $T_3$ intervention (see \circled{3} in \figref{fig:causal_settings_1}) explores the influence of model architecture on code smell propensity. We evaluated four language models with distinct architectural designs, denoted $M_1$ through $M_4$, and selected to be of comparable size (approximately 7B parameters). To control for generation variability, all models are evaluated using the same decoding strategy (\textit{greedy search}).

Following the same rationale as in the $T_2$ intervention, we treat each code fragment in the evaluation data set as a fixed input and compute the \psc ($Y_1 = f(\theta_\mu^{\text{median}} )$). This allows for a fair comparison of architectures under identical input conditions. 

In the causal setting, model $M_1$ serves as the control condition, denoted $T_3^0$. The remaining models ($M_2$, $M_3$, and $M_4$ are treated as intervention conditions, labeled $T_3^a$, $T_3^b$, and $T_3^c$, respectively. We estimate the average treatment effect by comparing the distribution of $Y_1$ across these architectural variants.

\subsubsection{Intervention Setting $P(Y_1 \mid do(T_4))$}

\begin{figure}[ht]
  \caption{Causal Analysis, $T_4$ Settings }
  \includegraphics[width=0.45\textwidth]{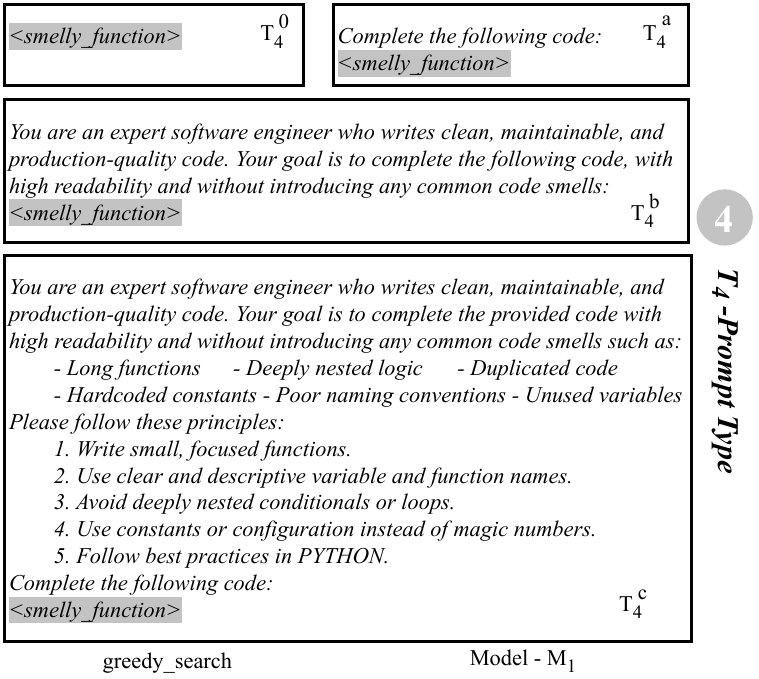}
    \label{fig:causal_settings_2}
\end{figure}

The $T_4$ intervention (see \circled{4} in \figref{fig:causal_settings_2}) investigates the influence of prompt type on code smell propensity. We define four prompt formulations that vary in their level of instruction and emphasis on code quality. The simplest version presents only a function placeholder (\texttt{<smelly\_function>}) with no additional context. The second variant appends a generic instruction: \textit{ 'Complete the following code'}. The third introduces a role-based preamble indicating that the user is an expert software engineer committed to producing clean, maintainable, and production-quality code. The fourth and most structured prompt explicitly lists code smells to avoid and includes best practices for modular and readable Python.

Similarly to interventions for $T_2$ and $T_3$, we isolate the effect of prompt design on \psc by evaluating all prompt variants using the same model ($M_1$) and a fixed decoding strategy (\textit{greedy search}). For each prompt condition, we compute the \psc for every snippet in the evaluation dataset, treating the snippet as the model output given the corresponding prompt.

In causal analysis, the minimal prompt (\texttt{<smelly\_function>}) is treated as the control condition, denoted $T_4^0$. The remaining three prompts are treated as intervention conditions: $T_4^a$, $T_4^b$, and $T_4^c$, reflecting increasing degrees of instructional guidance. We estimate the average treatment effects by comparing the resulting $Y_1$ distributions across prompt types.

\section{Results \& Discussion}
\label{sec:results_discussion}


\begin{table}[]
\centering
\caption{SEPT robustness \anova test results (\ie \ref{rq:measure}).}
\label{tab:robustness_results}
\scalebox{0.65}{

\setlength{\tabcolsep}{4pt} 
\begin{tabular}{lllllllllll}
\multicolumn{1}{c}{\textbf{ID}} &
  \multicolumn{1}{c}{\textbf{\anova}} &
  \multicolumn{1}{c}{\textbf{\pvalue}} &
  \multicolumn{1}{c}{\textbf{\etasquared}} &
  \textbf{\confidence} &
   &
  \multicolumn{1}{c}{\textbf{ID}} &
  \multicolumn{1}{c}{\textbf{\anova}} &
  \multicolumn{1}{c}{\textbf{\pvalue}} &
  \multicolumn{1}{c}{\textbf{\etasquared}} &
  \textbf{\confidence} \\ \cline{1-5} \cline{7-11} 
\textit{C0116} & 0       & 1 & 0      & 0.04 ± 0.003 &  & \textit{C3001}                         & 6.18e-2 & 1       & 1.2e-4  & 0.79 ± 0.009 \\
\textit{R0917} & 0       & 1 & 0      & 0.13 ± 0.007 &  & \textit{W1309}                         & 6.39e-2 & 1       & 1.3e-4  & 0.66 ± 0.012 \\
\textit{R1710} & 0       & 1 & 0      & 0.03 ± 0.003 &  & \textit{W0707}                         & 1.03e-1 & 9.81e-1 & 1.9e-4  & 0.81 ± 0.009 \\
\textit{R0914} & 0       & 1 & 0      & 0.03 ± 0.003 &  & \textit{W0612}                         & 1.25e-1 & 9.74e-1 & 2.8e-4  & 0.43 ± 0.014 \\
\textit{W0102} & 0       & 1 & 0      & 0.03 ± 0.003 &  & \textit{R1732}                         & 1.66e-1 & 9.56e-1 & 3.2e-4  & 0.77 ± 0.011 \\
\textit{C0114} & 0       & 1 & 0      & 0.24 ± 0.009 &  & \textit{W0311}                         & 2.76e-1 & 8.94e-1 & 5.2e-4  & 0.72 ± 0.011 \\
\textit{R0912} & 0       & 1 & 0      & 0.03 ± 0.003 &  & \textit{W0613}                         & 2.82e-1 & 8.9e-1  & 4.5e-4  & 0.27 ± 0.011 \\
\textit{R0913} & 0       & 1 & 0      & 0.12 ± 0.007 &  & \textit{C0301}                         & 0.36    & 0.83    & 6.3e-4  & 0.76 ± 0.011 \\
\textit{W0104} & 0       & 1 & 0      & 0.52 ± 0.017 &  & \textit{W0511}                         & 0.43    & 0.78    & 8.0e-4  & 0.37 ± 0.014 \\
\textit{R1735} & 1.98e-3 & 1 & 0      & 0.12 ± 0.007 &  & \textit{C0200}                         & 0.61    & 0.65    & 1.0e-3  & 0.96 ± 0.003 \\
\textit{W0622} & 2.72e-3 & 1 & 0      & 0.19 ± 0.011 &  & \textit{C0325}                         & 0.89    & 0.47    & 1.64e-3 & 0.75 ± 0.010 \\
\textit{C0415} & 7.34e-3 & 1 & 0      & 0.58 ± 0.012 &  & \textit{W0719}                         & 2.59    & 0.03    & 4.52e-3 & 0.72 ± 0.010 \\
\textit{C0303} & 8.72e-3 & 1 & 0      & 0.04 ± 0.006 &  & \textit{C0123}                         & 3.29    & 1.5e-2  & 0.13    & 0.71 ± 0.013 \\
\textit{W0601} & 1.02e-2 & 1 & 0      & 0.35 ± 0.011 &  & \textit{R1720}                         & 3.75    & 5e-3    & 6.35e-3 & 0.87 ± 0.005 \\
\textit{C0305} & 1.7e-2  & 1 & 3e-5   & 0.87 ± 0.009 &  & \cellcolor[HTML]{D4D4D4}\textit{R1705} & 3.93    & 3.49e-3 & 6.56e-3 & 0.88 ± 0.005 \\
\textit{C2801} & 1.95e-2 & 1 & 3e-5   & 0.81 ± 0.011 &  & \cellcolor[HTML]{D4D4D4}\textit{C0209} & 33.31   & 0       & 6.62e-2 & 0.88 ± 0.005 \\
\textit{C0304} & 2.44e-2 & 1 & 1.2e-3 & 0.51 ± 0.01  &  & \cellcolor[HTML]{D4D4D4}\textit{W0212} & 55.94   & 0       & 0.1     & 0.79 ± 0.011 \\
\textit{W0718} & 2.72e-2 & 1 & 4e-5   & 0.67 ± 0.012 &  & \cellcolor[HTML]{D4D4D4}\textit{C0321} & 65.85   & 0       & 0.12    & 0.72 ± 0.012 \\
\textit{W1406} & 3.61e-2 & 1 & 7e-5   & 0.74 ± 0.011 &  & \cellcolor[HTML]{D4D4D4}\textit{W1514} & 68.86   & 0       & 0.13    & 0.77 ± 0.012 \\
\textit{W0611} & 4.08e-2 & 1 & 7e-5   & 0.66 ± 0.012 &  & \cellcolor[HTML]{D4D4D4}\textit{C0103} & 94.51   & 0       & 0.18    & 0.32 ± 0.013 \\
\textit{W0621} & 5.41e-2 & 1 & 1.2e-4 & 0.39 ± 0.013 &  &                                        &         &         &         &             
\end{tabular}
} 
\caption*{\small{{\color[HTML]{c4c3c4} background grey}: \textnormal{reduced \psc robustness}}}
\end{table}

\begin{figure*}[ht]
		\centering
  \caption{Information Gain (\ig) Results (\ie \ref{rq:measure}).}
  \includegraphics[width=1\textwidth]{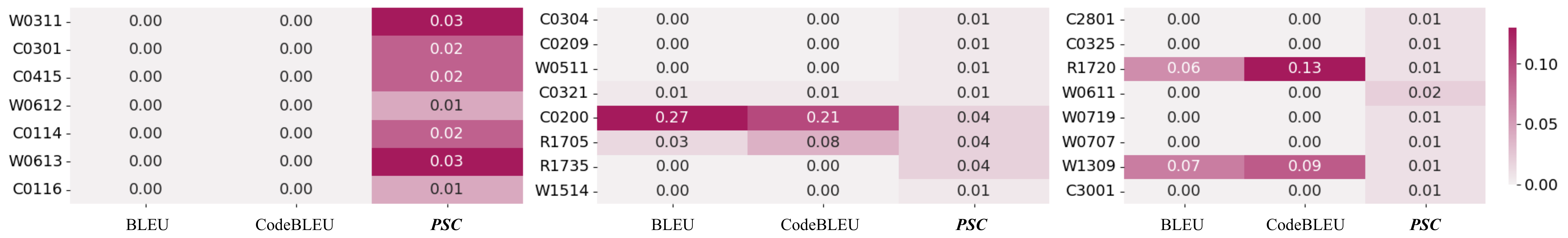}
    \label{fig:ig_results}
\end{figure*}

\tabref{tab:robustness_results} summarizes the results of our ANOVA analysis, which evaluates the robustness of the \psc metric under \sept. For $31$ out of $41$ code smell types (76\%), we observe $p$-values of 1 and effect sizes $\eta^2$ close to zero. This indicates that the distribution of the \psc scores remains consistent between syntactic variations that preserve the semantics of the code. These findings suggest that \psc captures stable model behavior and reflects deeper generative tendencies rather than shallow token patterns. Such stability is essential for a quality metric intended to evaluate the structural properties of code rather than incidental formatting.

However, a subset of smell types exhibits notable sensitivity. Specifically, \codesmell{C0103} (\textit{invalid}), \codesmell{C0209} (\textit{considering}), \codesmell{W0212} (\textit{protected}), and \codesmell{C0321} (\textit{multiple}) show higher effect sizes ($\eta^2 \geq 0.1$) and $p$-values below $.05$, with confidence intervals that vary more widely than in stable smells. These results, highlighted in gray in \tabref{tab:robustness_results}, indicate reduced robustness. In these cases, minor syntactic changes shift the model’s token likelihoods enough to affect the resulting \psc scores. Many of these smells depend on naming, formatting, or other localized syntactic patterns, which explains the increased variability in their confidence intervals. \revision{This pattern reflects a broader distinction between \textit{semantic} and \textit{syntactic} smells: semantic smells tend to remain stable under surface edits, whereas syntactic smells rely more directly on lexical and formatting characteristics and are therefore more sensitive to \sept.}

Across all types of smell, the average $std$ of the \psc scores was highest for \textit{SwitchRelation}, \textit{Add2Equal}, \textit{SwitchEqualExp}, and \textit{InfixDividing}, each averaging around $0.65$. In contrast, variability was lower for renaming-based transformations, with \textit{RenameVariable-1} averaging $0.44$ and \textit{RenameVariable-2} averaging $0.59$.

To assess the explanatory value of \psc relative to existing metrics, we computed the information gain (\ig) between each score (PSC, BLEU and CodeBLEU) and a binary label representing the severity of the smell of the code (see \secref{sec:ig}). The severity label assigns a positive class to snippets where more than $50$ percent of the tokens are marked as smelly. \figref{fig:ig_results} presents the results for all types of smells.

We find that \psc consistently produces a higher information gain than BLEU and CodeBLEU. This suggests that \psc is more effective in explaining the presence of severe code smells, offering a stronger and more discriminative signal about quality. Although BLEU and CodeBLEU focus on surface similarity with a reference implementation, they are less responsive to structural concerns such as maintainability or design violations. In contrast, \psc reflects the model's internal confidence in generating smelly tokens, making it more directly aligned with latent quality dimensions.

The results indicate that \psc offers two desirable properties for evaluating generated code: \textbf{robustness} to superficial edits and \textbf{alignment} with structural quality indicators. \psc demonstrates strong reliability under semantic-preserving transformations and provides more informative signals than reference-based metrics. Its sensitivity to certain smell types highlights the impact of local syntax on token prediction, but overall it offers a more interpretable and quality-aligned alternative for evaluating generated code.

\begin{boxK}
    \textit{\ref{rq:measure}} \textbf{[Measure]}: \psc provides a reliable way to measure the likelihood of code smell generation in \llm outputs. \psc remains robust under syntactic variation for 76\% of smell types and yields higher information gain than BLEU and CodeBLEU, indicating stronger alignment with smell severity and code quality.
\end{boxK}


\begin{table*}[]

\centering
\caption{Causal Effects (\ates) (\ie \ref{rq:explain})}
\label{tab:causal_results}
\scalebox{0.62}{

\setlength{\tabcolsep}{4pt} 
\begin{tabular}{lccccccccccllccccccllccccccllcccccc}
 &
  \multicolumn{10}{c}{\textit{Generation Type}} &
   &
   &
  \multicolumn{6}{c}{\textit{Model Size}} &
   &
  \multicolumn{1}{c}{\textit{}} &
  \multicolumn{6}{c}{\textit{Model Architecture}} &
  \multicolumn{1}{c}{\textit{}} &
  \multicolumn{1}{c}{\textit{}} &
  \multicolumn{6}{c}{\textit{Prompt Type}} \\
 &
  \multicolumn{2}{c}{$T_1^a$} &
  \multicolumn{2}{c}{$T_1^b$} &
  \multicolumn{2}{c}{$T_1^c$} &
  \multicolumn{2}{c}{$T_1^d$} &
  \multicolumn{2}{c}{$T_1^e$} &
   &
   &
  \multicolumn{2}{c}{$T_2^a$} &
  \multicolumn{2}{c}{$T_2^b$} &
  \multicolumn{2}{c}{$T_2^b$} &
   &
  \multicolumn{1}{c}{} &
  \multicolumn{2}{c}{$T_3^a$} &
  \multicolumn{2}{c}{$T_3^b$} &
  \multicolumn{2}{c}{$T_3^c$} &
  \multicolumn{1}{c}{} &
  \multicolumn{1}{c}{} &
  \multicolumn{2}{c}{$T_4^a$} &
  \multicolumn{2}{c}{$T_4^b$} &
  \multicolumn{2}{c}{$T_4^c$} \\
\textbf{ID} &
  \textbf{\pearson} &
  \textbf{\ate} &
  \textbf{\pearson} &
  \textbf{\ate} &
  \textbf{\pearson} &
  \textbf{\ate} &
  \textbf{\pearson} &
  \textbf{\ate} &
  \textbf{\pearson} &
  \textbf{\ate} &
  \textbf{} &
  \textbf{ID} &
  \textbf{\pearson} &
  \textbf{\ate} &
  \textbf{\pearson} &
  \textbf{\ate} &
  \textbf{\pearson} &
  \textbf{\ate} &
   &
  \textbf{ID} &
  \textbf{\pearson} &
  \textbf{\ate} &
  \textbf{\pearson} &
  \textbf{\ate} &
  \textbf{\pearson} &
  \textbf{\ate} &
   &
  \textbf{ID} &
  \textbf{\pearson} &
  \textbf{\ate} &
  \textbf{\pearson} &
  \textbf{\ate} &
  \textbf{\pearson} &
  \textbf{\ate} \\ \cline{1-11} \cline{13-19} \cline{21-27} \cline{29-35} 
\textit{C0303} &
  {\ul \textbf{.38}} &
  \cellcolor[HTML]{D4D4D4}.15 &
  {\ul \textbf{.44}} &
  \cellcolor[HTML]{D4D4D4}.21 &
  {\ul \textbf{.51}} &
  \cellcolor[HTML]{D4D4D4}.23 &
  {\ul \textbf{.52}} &
  \cellcolor[HTML]{D4D4D4}.28 &
  {\ul \textbf{.51}} &
  \cellcolor[HTML]{D4D4D4}.26 &
   &
  \textit{C0103} &
  .04 &
  \cellcolor[HTML]{D4D4D4}.10 &
  .06 &
  \cellcolor[HTML]{D4D4D4}.17 &
  .08 &
  -.02 &
   &
  C0103 &
  \textbf{-.10} &
  \cellcolor[HTML]{D4D4D4}.24 &
  {\ul \textbf{.10}} &
  .04 &
  {\ul \textbf{.17}} &
  \cellcolor[HTML]{D4D4D4}32 &
   &
  \textit{C0209} &
  -.01 &
  -.05 &
  -.03 &
  -.06 &
  -.03 &
  \cellcolor[HTML]{C4739C}-.10 \\
\textit{C0304} &
  {\ul \textbf{.27}} &
  \cellcolor[HTML]{D4D4D4}.1.0 &
  {\ul \textbf{.55}} &
  \cellcolor[HTML]{D4D4D4}.30 &
  {\ul \textbf{.32}} &
  \cellcolor[HTML]{D4D4D4}.17 &
  {\ul \textbf{.24}} &
  \cellcolor[HTML]{D4D4D4}.15 &
  {\ul \textbf{.31}} &
  \cellcolor[HTML]{D4D4D4}.17 &
   &
  \textit{C0200} &
  {\ul \textbf{.20}} &
  \cellcolor[HTML]{D4D4D4}.10 &
  {\ul \textbf{.25}} &
  \cellcolor[HTML]{D4D4D4}.11 &
  {\ul \textbf{.34}} &
  \cellcolor[HTML]{D4D4D4}.14 &
   &
  \textit{C0114} &
  \textbf{-.23} &
  \cellcolor[HTML]{C4739C}-.14 &
  \textbf{-.31} &
  \cellcolor[HTML]{C4739C}-.14 &
  .08 &
  .06 &
   &
  \textit{C0304} &
  -.07 &
  \cellcolor[HTML]{C4739C}-.50 &
  \textbf{-.10} &
  \cellcolor[HTML]{C4739C}-.52 &
  \textbf{-.10} &
  \cellcolor[HTML]{C4739C}-.52 \\
\textit{C2801} &
  -.03 &
  0.0 &
  {\ul \textbf{.12}} &
  -.01 &
  .04 &
  \cellcolor[HTML]{C4739C}-.14 &
  .02 &
  -.04 &
  .07 &
  \cellcolor[HTML]{C4739C}-.10 &
   &
  \textit{C0209} &
  .08 &
  .05 &
  {\ul \textbf{.10}} &
  \cellcolor[HTML]{D4D4D4}.12 &
  {\ul \textbf{.16}} &
  \cellcolor[HTML]{D4D4D4}.15 &
   &
  \textit{C0304} &
  \textbf{-.14} &
  \cellcolor[HTML]{C4739C}-.53 &
  \textbf{-.15} &
  \cellcolor[HTML]{C4739C}-.25 &
  \textbf{-.22} &
  \cellcolor[HTML]{C4739C}-.55 &
   &
  \textit{C0321} &
  0.0 &
  \cellcolor[HTML]{D4D4D4}.13 &
  -.01 &
  \cellcolor[HTML]{D4D4D4}.15 &
  -.01 &
  \cellcolor[HTML]{D4D4D4}.10 \\
\textit{R1732} &
  \textbf{-.11} &
  0.0 &
  -.01 &
  -.06 &
  -.09 &
  \cellcolor[HTML]{C4739C}-.12 &
  -.07 &
  -.06 &
  -.04 &
  .05 &
   &
  \textit{C0301} &
  {\ul \textbf{.13}} &
  \cellcolor[HTML]{D4D4D4}.12 &
  {\ul \textbf{.16}} &
  \cellcolor[HTML]{D4D4D4}.25 &
  {\ul \textbf{.23}} &
  \cellcolor[HTML]{D4D4D4}.30 &
   &
  \textit{C0305} &
  -.07 &
  -.03 &
  -.04 &
  -.02 &
  \textbf{-.40} &
  \cellcolor[HTML]{C4739C}-.26 &
   &
  \textit{C0325} &
  -.02 &
  -.01 &
  -.04 &
  \cellcolor[HTML]{C4739C}-.12 &
  -.04 &
  \cellcolor[HTML]{C4739C}-.11 \\
\textit{R1735} &
  -.04 &
  -.03 &
  {\ul \textbf{.10}} &
  -.03 &
  0.0 &
  -.08 &
  -.01 &
  \cellcolor[HTML]{C4739C}-.10 &
  .03 &
  .09 &
   &
  \textit{C0304} &
  .06 &
  .01 &
  .08 &
  \cellcolor[HTML]{D4D4D4}.16 &
  {\ul \textbf{.14}} &
  \cellcolor[HTML]{D4D4D4}.33 &
   &
  C0321 &
  -.06 &
  .06 &
  \textbf{-.11} &
  \cellcolor[HTML]{D4D4D4}.15 &
  -.02 &
  \cellcolor[HTML]{D4D4D4}.18 &
   &
  \textit{R0917} &
  -.05 &
  -.04 &
  -.06 &
  -.05 &
  \textbf{-.18} &
  \cellcolor[HTML]{C4739C}-.13 \\
\textit{W0104} &
  {\ul \textbf{.10}} &
  \cellcolor[HTML]{D4D4D4}.1.0 &
  {\ul \textbf{.33}} &
  .06 &
  {\ul \textbf{.16}} &
  .04 &
  .09 &
  .01 &
  {\ul \textbf{.22}} &
  \cellcolor[HTML]{D4D4D4}.13 &
   &
  \textit{C0321} &
  .09 &
  \cellcolor[HTML]{D4D4D4}.28 &
  {\ul \textbf{.11}} &
  \cellcolor[HTML]{D4D4D4}.39 &
  {\ul \textbf{.16}} &
  \cellcolor[HTML]{D4D4D4}.42 &
   &
  \textit{C0415} &
  \textbf{-.18} &
  \cellcolor[HTML]{C4739C}-.16 &
  \textbf{-.16} &
  \cellcolor[HTML]{C4739C}-.38 &
  \textbf{-.18} &
  \cellcolor[HTML]{C4739C}-.33 &
   &
  \textit{W0102} &
  .08 &
  .06 &
  .09 &
  .06 &
  {\ul \textbf{.22}} &
  \cellcolor[HTML]{D4D4D4}.19 \\
\textit{W0511} &
  -.07 &
  -.07 &
  {\ul \textbf{.35}} &
  \cellcolor[HTML]{D4D4D4}.24 &
  .02 &
  -.07 &
  .04 &
  0.0 &
  .03 &
  -.04 &
   &
  \textit{C0325} &
  {\ul \textbf{.10}} &
  .04 &
  {\ul \textbf{.13}} &
  \cellcolor[HTML]{D4D4D4}.13 &
  {\ul \textbf{.18}} &
  \cellcolor[HTML]{D4D4D4}.16 &
   &
  \textit{C3001} &
  \textbf{-.15} &
  \cellcolor[HTML]{C4739C}-.19 &
  \textbf{-.13} &
  \cellcolor[HTML]{C4739C}-.24 &
  \textbf{-.10} &
  -.07 &
   &
  \textit{W0601} &
  -.08 &
  -.06 &
  \textbf{-.12} &
  -.06 &
  \textbf{-.10} &
  \cellcolor[HTML]{D4D4D4}.11 \\
\textit{W0612} &
  .03 &
  \cellcolor[HTML]{D4D4D4}.12 &
  {\ul \textbf{.25}} &
  \cellcolor[HTML]{D4D4D4}.13 &
  {\ul \textbf{.19}} &
  \cellcolor[HTML]{D4D4D4}.21 &
  {\ul \textbf{.13}} &
  \cellcolor[HTML]{D4D4D4}.19 &
  {\ul \textbf{.20}} &
  \cellcolor[HTML]{D4D4D4}.19 &
   &
  \textit{W0511} &
  .05 &
  .07 &
  .06 &
  \cellcolor[HTML]{D4D4D4}.12 &
  .09 &
  \cellcolor[HTML]{D4D4D4}.14 &
   &
  \textit{R0917} &
  \textbf{-.27} &
  \cellcolor[HTML]{C4C3C4}-.19 &
  .06 &
  .08 &
  {\ul \textbf{.30}} &
  \cellcolor[HTML]{D4D4D4}.30 &
   &
  \textit{W0611} &
  -.04 &
  \cellcolor[HTML]{C4739C}-.12 &
  -.08 &
  \cellcolor[HTML]{C4739C}-.12 &
  -.07 &
  \cellcolor[HTML]{C4739C}-.11 \\
\textit{W0613} &
  .08 &
  .05 &
  {\ul \textbf{.13}} &
  -.01 &
  {\ul \textbf{.25}} &
  \cellcolor[HTML]{D4D4D4}.16 &
  {\ul \textbf{.16}} &
  \cellcolor[HTML]{D4D4D4}.19 &
  {\ul \textbf{.30}} &
  \cellcolor[HTML]{D4D4D4}.19 &
   &
  \textit{W0613} &
  .06 &
  \cellcolor[HTML]{D4D4D4}.21 &
  {\ul \textbf{.13}} &
  \cellcolor[HTML]{D4D4D4}.33 &
  {\ul \textbf{.10}} &
  \cellcolor[HTML]{D4D4D4}.31 &
   &
  \textit{R1705} &
  \textbf{-.18} &
  -.05 &
  \textbf{-.35} &
  \cellcolor[HTML]{C4739C}-.17 &
  \textbf{-.25} &
  \cellcolor[HTML]{C4739C}-.15 &
   &
  \textit{W0707} &
  0.0 &
  -.01 &
  -.02 &
  \cellcolor[HTML]{C4739C}-.13 &
  -.01 &
  \cellcolor[HTML]{C4739C}-.12 \\
\textit{W0621} &
  -.05 &
  0.0 &
  {\ul \textbf{.12}} &
  \cellcolor[HTML]{D4D4D4}.10 &
  .07 &
  \cellcolor[HTML]{D4D4D4}.12 &
  .07 &
  \cellcolor[HTML]{D4D4D4}.14 &
  .07 &
  \cellcolor[HTML]{D4D4D4}.12 &
   &
  \textit{W1514} &
  .05 &
  \cellcolor[HTML]{D4D4D4}.11 &
  .04 &
  \cellcolor[HTML]{D4D4D4}.10 &
  .08 &
  \cellcolor[HTML]{D4D4D4}.13 &
   &
  \textit{R1720} &
  \textbf{-.15} &
  -.03 &
  \textbf{-.35} &
  \cellcolor[HTML]{C4739C}-.16 &
  \textbf{-.28} &
  \cellcolor[HTML]{C4739C}-.11 &
   &
  \textit{W0719} &
  -.02 &
  -.05 &
  -.01 &
  \cellcolor[HTML]{C4739C}-.13 &
  -.02 &
  \cellcolor[HTML]{C4739C}-.15 \\
\textit{W1309} &
  \textbf{-.35} &
  -.05 &
  .07 &
  \cellcolor[HTML]{C4739C}-.10 &
  .02 &
  -.02 &
  -.09 &
  \cellcolor[HTML]{C4739C}-.12 &
  0.0 &
  \cellcolor[HTML]{C4739C}-.17 &
   &
   &
   &
   &
   &
   &
   &
   &
   &
  \textit{W0601} &
  \textbf{-.20} &
  \cellcolor[HTML]{C4739C}-.23 &
  \textbf{-.15} &
  \cellcolor[HTML]{C4739C}-.14 &
  \textbf{-1.0} &
  \cellcolor[HTML]{C4739C}-.23 &
   &
   &
   &
   &
   &
   &
   &
   \\
\textit{W1406} &
  {\ul \textbf{.20}} &
  \cellcolor[HTML]{D4D4D4}.15 &
  {\ul \textbf{.20}} &
  0.0 &
  {\ul \textbf{.12}} &
  .03 &
  {\ul \textbf{.13}} &
  \cellcolor[HTML]{D4D4D4}.40 &
  {\ul \textbf{.16}} &
  \cellcolor[HTML]{C4739C}-.16 &
   &
   &
   &
   &
   &
   &
   &
   &
   &
  \textit{W0612} &
  \textbf{-.20} &
  \cellcolor[HTML]{C4739C}-.22 &
  \textbf{-.15} &
  \cellcolor[HTML]{C4739C}-.14 &
  \textbf{-1.0} &
  -.07 &
   &
   &
   &
   &
   &
   &
   &
   \\
 &
   &
   &
   &
   &
   &
   &
   &
   &
   &
   &
   &
   &
   &
   &
   &
   &
   &
   &
   &
  \textit{W0707} &
  \textbf{-.12} &
  \cellcolor[HTML]{C4739C}-.17 &
  \textbf{-.29} &
  \cellcolor[HTML]{C4739C}-.39 &
  \textbf{-.32} &
  \cellcolor[HTML]{C4739C}-.34 &
   &
   &
   &
   &
   &
   &
   &
   \\
 &
   &
   &
   &
   &
   &
   &
   &
   &
   &
   &
   &
   &
   &
   &
   &
   &
   &
   &
   &
  \textit{W1406} &
  -.06 &
  -.09 &
  \textbf{-.18} &
  \cellcolor[HTML]{C4739C}-.27 &
  \textbf{-.11} &
  \cellcolor[HTML]{C4739C}-.14 &
   &
   &
   &
   &
   &
   &
   &
   \\
 &
   &
   &
   &
   &
   &
   &
   &
   &
   &
   &
   &
   &
   &
   &
   &
   &
   &
   &
   &
  \textit{W0719} &
  -.09 &
  \cellcolor[HTML]{C4739C}-.17 &
  \textbf{-.26} &
  \cellcolor[HTML]{C4739C}-.10 &
  \textbf{-.19} &
  -.06 &
   &
   &
   &
   &
   &
   &
   &
  
\end{tabular}
} 
\caption*{\small{bold: \textnormal{$-$ correlation}, \underline{bold underlined}: \textnormal{$+$ correlation}, {\color[HTML]{c4739c} background purple}: \textnormal{$-$ causal effect}, {\color[HTML]{c4c3c4} background grey}: \textnormal{$+$ causal effect}}}
\end{table*}

\tabref{tab:causal_results} reports the average treatment effects (\ates) estimated across four causal interventions (\secref{sec:causal_settings}). Results indicate varying sensitivity across code smell types, with some interventions producing stronger effects than others.

\textit{Generation type ($T_1$)} leads to substantial variation in \psc scores depending on the decoding strategy. Smells such as \codesmell{C2801} (\textit{unnecessary -dunder-call}) and \codesmell{W1406} (\textit{redundant-u-string-prefix}) show lower \psc values when sampling-based methods like $T_1^c$ or $T_1^e$ are used, suggesting reduced smell propensity under stochastic generation. Conversely, formatting-related smells, including \codesmell{C0303} (\textit{trailing-whitespace}) and \codesmell{C0304} (\textit{missing-final-newline}), tend to increase under non-greedy decoding, reflecting greater variability in surface-level code structure.

\textit{Model size ($T_2$)} exhibits relatively weak influence on the \psc. Most treatment effects remain close to zero, and even slight improvements observed for convention-type smells lack statistical significance. These findings suggest that scaling the number of parameters alone does not substantially reduce the likelihood of generating smelly code, at least within the model sizes tested. The assumption that larger models automatically produce cleaner outputs is not supported.

\textit{Model architecture ($T_3$)} has a more pronounced impact. Several refactor and warning-related smells, including \codesmell{R1705} (\textit{no-else-return}), \codesmell{W0612} (\textit{unused-variable}), and \codesmell{W0707} (\textit{raise-missing-from}), display large negative treatment effects across architectural variants. Changes in network design influence the model’s internal token distributions, affecting structural and semantic decisions during code synthesis. Even stylistic issues, such as \codesmell{C0304} and \codesmell{C0325}, are affected, reinforcing the role of architecture in shaping code quality.

\textit{Prompt design ($T_4$)} consistently reduces smell propensity, particularly when instructions emphasize best practices and explicitly discourage problematic patterns. Strongest effects are observed for \codesmell{W0612}, \codesmell{W0707}, and \codesmell{C0321}, where more structured prompts yield noticeable reductions in \psc scores. Prompt variant $T_4^c$, which includes specific smell avoidance guidance, produces the most substantial improvement. Instructional framing thus emerges as an effective lever for steering generation without retraining the model.

Model architecture and prompt formulation emerge as effective levers for mitigating code smell generation, while model size offers minimal benefit. Decoding strategy shows mixed effects, with improvements for some semantic issues and degradation for surface-level ones. Prompt engineering, in particular, offers a practical intervention for developers, requiring no changes to the model itself. Architectural findings suggest deeper interactions between pretraining design and output quality, opening avenues for more architecture-aware alignment strategies.


Refutation tests confirm the stability of the estimated \ates. Results remain consistent across validation checks, indicating robustness to noise and unmeasured factors.

\begin{boxK}
    \textit{\ref{rq:explain}} \textbf{[Explain]}: The \ates reveal that the \psc is differentially influenced by generation type, model size, architecture, and prompt design. Although model size shows limited impact, architectural changes and prompt formulation exhibit strong and consistent effects on reducing the propensity for specific code smells.
\end{boxK}



\begin{figure}[ht]
		\centering
  \caption{Mitigation Case Study Results (\ie \ref{rq:mitigate}).}
  \includegraphics[width=0.47\textwidth]{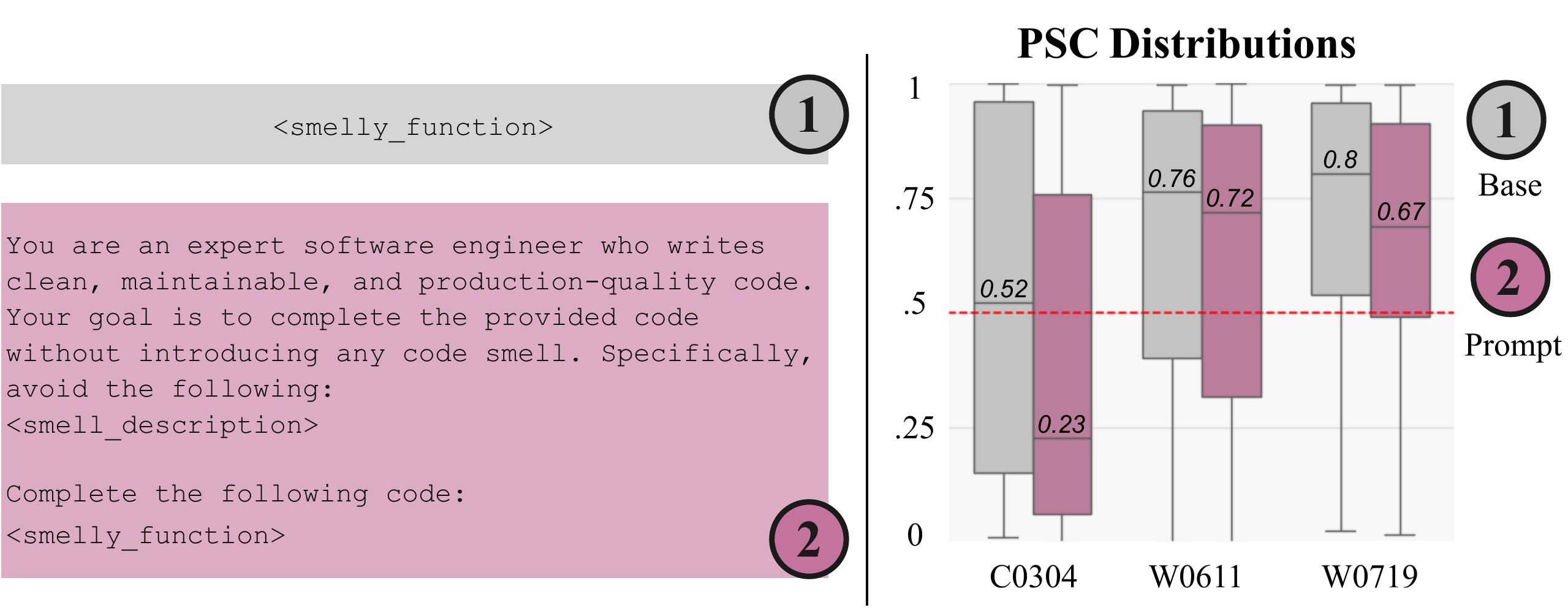}
    \label{fig:mitigation_results}
\end{figure}



\revision{\figref{fig:mitigation_results} shows the impact of prompt-based mitigation on the \psc score for three smell types: \codesmell{W0719} (\textit{broad-exception-raised}), \codesmell{C0304} (\textit{missing-final-newline}) and \codesmell{W0611} (\textit{unused-import}). The prompt instructs the model to complete the provided code without introducing any of the listed smells. To assess its effect, we constructed a new evaluation set of $500$ examples for each smell type. All generations were produced using model $M_1$ with greedy decoding under two conditions: a baseline prompt containing only the original function (\circled{1}) and a mitigation prompt including the instruction (\circled{2}).}

\revision{The boxplots in \figref{fig:mitigation_results} compare the resulting \psc distributions across both conditions. For \codesmell{W0719}, the median \psc decreases from $.80$ under the baseline to $.67$ with mitigation. For \codesmell{C0304}, the median decreases from $.76$ to $.72$, and for \codesmell{W0611} the reduction is more pronounced, from $.52$ to $.23$. In all three cases, the distributions shift downward, with a larger proportion of instances falling below the smelly threshold $\lambda = 0.5$.}

\revision{
Because this intervention relies exclusively on prompt design, it requires no additional training cost and applies directly at inference time. The consistent declines across the three smell types suggest that prompt-based mitigation is an effective and scalable strategy for improving the quality of \llm-generated code.
}

\begin{boxK}
\revision{\textit{\ref{rq:mitigate}} \textbf{[Mitigate]}: Prompt-based mitigation reduces the median \psc for \codesmell{W0719} from $.80$ to $.67$, for \codesmell{C0304} from $.76$ to $.72$, and for \codesmell{W0611} from $.52$ to $.23$. These results show that carefully crafted prompts reduce the model’s propensity to generate code smells during inference.}
\end{boxK}



We assessed whether the \psc score influences how developers evaluate code smells through a between-subjects user study comparing two conditions: a control group that viewed code samples without the \psc score, and a treatment group that received the same samples with the \psc score displayed. \tabref{tab:survey_results} summarizes the statistical results, with significant $p$-values ($p < .05$) highlighted in purple.

Five significant effects were observed across three smell types. For \codesmell{W0613} (\textit{unused-argument}), access to the \psc score increased the perceived importance of fixing the smell ($Q_2$, $p = .034$) and boosted participants’ confidence in their judgment ($Q_3$, $p = .002$). For \codesmell{W0612} (\textit{unused-variable}), participants in the treatment group were more likely to attribute the smell to systematic model behavior ($Q_1$, $p = .002$). In the case of \codesmell{W0622} (\textit{redefined-builtin}), participants exposed to the score more frequently judged the smell as systematically introduced ($Q_1$, $p = .026$) and also rated it as more important to address ($Q_2$, $p = .021$).

Responses to $Q_4$ suggest that the \psc score helped participants reassess their initial judgments, especially when the smell was subtle or unfamiliar. Some used it as a reference point to confirm concerns or adjust prioritization, with high scores increasing urgency and low scores offering reassurance. Overall, the score functioned as a heuristic for resolving uncertainty in edge cases.  

The \psc score influenced how developers interpreted and prioritized code smells, especially for subtle issues like variable usage and naming. Its impact varied by smell type, suggesting that \psc is most useful in ambiguous cases rather than as a universal signal.

\begin{table}[]
\centering
\caption{User Study Results (\ie \ref{rq:usefulness}).}
\label{tab:survey_results}
\scalebox{0.8}{
\setlength{\tabcolsep}{4pt} 
\begin{tabular}{lllllllll}
\textbf{ID} &
  \textbf{$Q_{ID}$} &
  \textbf{\textit{U-stat}} &
  \textbf{\pvalue} &
  \textbf{} &
  \textbf{ID} &
  \textbf{$Q_{ID}$} &
  \textbf{\textit{U-stat}} &
  \textbf{\pvalue} \\ \cline{1-4} \cline{6-9} 
C0209 &
  1 &
  138 &
  .437 &
   &
  W0102 &
  1 &
  186.5 &
  .429 \\
C0209 &
  2 &
  154 &
  .726 &
   &
  W0102 &
  2 &
  130 &
  .863 \\
C0209 &
  3 &
  136 &
  .393 &
   &
  W0102 &
  3 &
  172 &
  .751 \\
W0613 &
  1 &
  162 &
  1 &
   &
  W0622 &
  1 &
  230 &
  .026 \\
\cellcolor[HTML]{C4739C}W0613 &
  \cellcolor[HTML]{C4739C}2 &
  \cellcolor[HTML]{C4739C}84.5 &
  \cellcolor[HTML]{C4739C}.034 &
   &
  \cellcolor[HTML]{C4739C}W0622 &
  \cellcolor[HTML]{C4739C}2 &
  \cellcolor[HTML]{C4739C}78.5 &
  \cellcolor[HTML]{C4739C}.021 \\
\cellcolor[HTML]{C4739C}W0613 &
  \cellcolor[HTML]{C4739C}3 &
  \cellcolor[HTML]{C4739C}67.5 &
  \cellcolor[HTML]{C4739C}.002 &
   &
  W0622 &
  3 &
  167 &
  .881 \\
\cellcolor[HTML]{C4739C}W0612 &
  \cellcolor[HTML]{C4739C}1 &
  \cellcolor[HTML]{C4739C}89 &
  \cellcolor[HTML]{C4739C}.002 &
   &
   &
   &
   &
   \\
W0612 &
  2 &
  113.5 &
  .422 &
   &
   &
   &
   &
   \\
W0612 &
  3 &
  153 &
  .771 &
   &
   &
   &
   &
  
\end{tabular}
} 
\caption*{\small{{\color[HTML]{c4739c} background purple}: \textnormal{$-$ statistically significant}}}
\end{table}

\begin{boxK}
    \textit{\ref{rq:usefulness}} \textbf{[Usefulness]}: Exposure to the \psc helped participants identify smells as model-introduced, prioritize them for resolution, and evaluate them with greater confidence. The perceived usefulness of \psc varied by smell type.
\end{boxK}


\section{Threats to Validity}
\label{sec:threats_to_validity}

Threats to \textbf{internal validity} concern the correctness of our implementation, causal reasoning, and study design. To mitigate these, we leverage established tools such as \texttt{DoWhy}\cite{dowhy} for causal inference and carefully validate our \psc implementation. We perform multiple robustness checks, including placebo tests, randomized confounding, and subset validation, to ensure the reliability of our causal estimates. In the user study, we controlled for potential ordering effects and standardized question phrasing to minimize participant bias.

\textbf{External validity} relates to the extent to which our findings generalize beyond the specific settings of our study. Although our experiments focus on Python and decoder-only language models, we incorporate diverse architectures and analyze 41 distinct Pylint smell types (\tabref{tab:robustness_results}), which enhances representativeness within our chosen domain. Generalization to other programming languages or tooling ecosystems remains an avenue for future work.
%


\textbf{Construct validity} addresses how accurately our metrics capture the underlying concept of code quality. We quantify smell propensity through token-level smell density and estimate generation likelihood using \psc, grounded in static analysis techniques. The alignment between these metrics and perceived quality is supported by robustness analysis and user study feedback. While \psc builds on established definitions of code smells and reflects structural indicators of quality, \revision{our causal analysis depends on a confounder set drawn from features most commonly associated with smell occurrence. Other design attributes such as member hierarchy, graph call patterns, or maintainability indices were not included because they cannot be reliably extracted from isolated snippets. These omissions may introduce residual confounding, although refutation tests suggest that our estimates remain stable. Future work could incorporate richer design-level signals to broaden the confounder set.}

\section{Conclusions \& Future Work}
\label{sec:conclusions_future_work}

In this paper, we used \psc to measure, explain, and mitigate the generation of code smells in \llms. Our findings show that \psc aligns more closely with structural quality than surface-level metrics such as BLEU and CodeBLEU, and remains robust under semantic-preserving transformations. Causal analysis revealed that prompt design and model architecture are key factors that influence the propensity to smell, while the size of the model has limited impact. We also demonstrated that prompt-based mitigation can significantly reduce the likelihood of generating smelly code. A user study confirmed that \psc improves developer judgement in ambiguous scenarios.

\revision{Future work includes extending \psc to other programming languages and model architectures, which will require adapting smell taxonomies and ensuring access to next-token probabilities. Another direction is the development of full mitigation plans that move beyond the case study explored in this work. More broadly, \psc has potential as a lightweight interpretability signal beyond code quality, offering a practical way to examine how model probability distributions behave and what these patterns reveal about underlying generation tendencies.}

\section{Acknowledgments}
This research has been supported in part by the NSF
CCF-2311469 and CCF-2346357. We also gratefully acknowledge support from Cisco Systems. 
The opinions, findings and conclusions
expressed in this work are solely those of the authors and do
not necessarily reflect the views of the sponsors.

\balance
\bibliographystyle{ACM-Reference-Format}
\bibliography{main.bib}

\end{document}